\documentclass[fleqn,usenatbib]{mnras}
\DeclareSymbolFont{cmletters}{OML}{cmm}{m}{it}
\DeclareMathSymbol{v}{\mathalpha}{cmletters}{"76}

\usepackage{ae,aecompl}
\usepackage{times}
\usepackage{soul}

\usepackage{tabularx,ragged2e,booktabs,caption}
\usepackage{amsmath}
\usepackage{amssymb}
\usepackage{epsfig}
\usepackage{graphicx}
\usepackage{ifthen}
\usepackage{latexsym}
\usepackage{rotating}
\usepackage{times,epsf}
\usepackage{txfonts}
\usepackage{varioref}
\usepackage{verbatim}
\usepackage{array}
\usepackage{url}
\usepackage{color}
\usepackage[T1]{fontenc}
\usepackage{epstopdf}
\usepackage{ulem}

\def\gsim{\mathrel{\raise.5ex\hbox{$>$}\mkern-14mu
             \lower0.6ex\hbox{$\,\sim$}}}
\def\lsim{\mathrel{\raise.3ex\hbox{$<$}\mkern-14mu
             \lower0.6ex\hbox{$\,\sim$}}}

\newcommand{\be}{\begin{equation}}
\newcommand{\ee}{\end{equation}}
\newcommand{\bea}{\begin{eqnarray}}
\newcommand{\eea}{\end{eqnarray}}

\title{Gravitational Waves from GRB Core Spindown}
\author[I. Contopoulos]
       {I. Contopoulos$^{1}$\thanks{E-mail: icontop@academyofathens.gr}, A. Strantzalis$^{2}$, D. Papadopoulos$^{3}$, D. Kazanas$^{4}$\\
$^1$ Research Center for Astronomy and Applied Mathematics, Academy of Athens, Athens 11527, Greece\\
$^2$ Department of Physics, National and Kapodistrian University of Athens, Panepistimiopolis, Zografos, GR15784, Greece\\
$^3$ Aristotle University of Thessaloniki, Department of Physics, University Campus Laboratory of Astronomy, 54124 Thessaloniki, Greece\\
$^4$ NASA/Goddard Space Flight Center, Greenbelt, MD 20771, USA
}

\begin{document}

\maketitle

\label{firstpage}

\begin{abstract}
We investigate long Gamma-Ray Bursts (GRB) which manifest a sharp linear rise followed by an exponential decay in their $\gamma$-ray prompt emission observed with the BAT instrument on board the Swift satellite. We offer a simple electrodynamic model that may account for these particular characteristics. We associate the sharp rise with the winding of the magnetic field by the fast rotating core that formed in the interior of the stellar precursor. We also associate the subsequent exponential decay with the electromagnetic spindown of the core following the release of the electromagnetic jet from the stellar interior. Any non-axisymmetric distortion in the rotating core will generate gravitational waves with exponentially decreasing frequency, a so-called `down-chirp'. We obtain a detailed estimate of the gravitational wave profile if the distortion of spacetime is due to the winding of a non-axisymmetric component of the magnetic field during that particular phase of the burst. We offer 7 particular time intervals during which one may look into LIGO archival data for the presence of our particular predicted waveforms in order to test our interpretation.
\end{abstract}


\section{Introduction}

Long duration Gamma-Ray Bursts (hereafter GRB) have long been associated with the gravitatonal collapse of the cores of supermassive stars and the subsequent formation of collimated magnetically driven relativistic outflows/jets \citep{WB06}. The process may seem simple, but the resulting prompt emission X-ray and $\gamma$-ray light curves vary dramatically in morphology.

In this work, we are looking for GRB prompt emission with particular temporal characteristics namely a fast linear rise followed by a slower exponential decay. The reason is that, for these particular light curves, we can formulate a simple model for the electrodynamic spindown of the newly formed core. Furthermore, under certain circumstances, the spinning down core will generate Gravitational Waves (hereafter GW) that may be detectable with the LIGO GW observatory. In fact, our simple model predicts a particular GW waveform {\it with decreasing frequency} that may be retrievable from LIGO archival GW data during the time intervals of the particular prompt emission lightcurves.

In the next session we present our analysis of archival BAT observations that resulted in a Table of 28 events with the particular characteristics of a fast linear rise followed by a longer exponential decay. In \S~3 we present a simple model that describes the electromagnetic spindown of the newly formed collapsed core and results in the generation of the observed GRB prompt emission light curves of the Table. In \S~4 we suggest that all of these events are potential sources of GW with a particular waveform which we calculate. We conclude this work by urging the LIGO Team to look into their archival data for these particular GW waveforms during 7 particular time intervals that we specify. This will confirm or disprove our simple model.

\section{Prompt emission observations}

We searched the BAT/Swift database of prompt emission lightcurves in the 15-50 keV band\footnote{https://www.swift.ac.uk/burst\_analyser/  \citep{Letal16}.} 
and looked for lightcurves with the following particular characteristics: a sharp linear rise followed by a longer exponential decay. For each lightcurve, we fit the following profile:
\begin{eqnarray}
\dot{E}_{\rm 15-50\ keV}(t) 
& = & \left\{\begin{array}{lr}
        at+b & \text{for  } 0\leq t\leq \Delta\\ \\
        A{\rm e}^{-\frac{t-\Delta}{\tau}} & \text{for  } t> \Delta
        \end{array}\right. \nonumber\\
        \label{fit}
\end{eqnarray}
We show a characteristic example in figure~1 where we also show the coefficient of determination ($R^2$) of our fit. This investigation resulted in the list of 28 lightcurves shown in Table~1. The purpose of this effort is to formulate a simple physical model that accounts for these features from which we can make predictions about the physical parameters at the origin of long duration GRBs. Notice that \citet{Netal05} has proposed the phenomenological fit $\dot{E}_{\rm GRB}\propto {\rm e}^{-\frac{t-\Delta}{\tau_2}-\frac{\tau_1}{t-\Delta}}$ for both the increasing and decreasing part of the burst in the prompt emission. Their fit refers to GRB events observed with BATSE and there is no specific mention of an early linear rise in them. At late times the two fits are similar. However, their correponding exponential decay timescales $\tau$ ($\tau_2$ in their Table~2) are systematically longer than the timescales we obtain for the events in our Table~1 below. As we will see in the next section, our fit is not only simpler, but is also amenable to a simple testable physical interpretation. 
\begin{figure}
 \centering
 %\vspace{3cm}
 \includegraphics[width=9cm,height=7cm,angle=0.0]{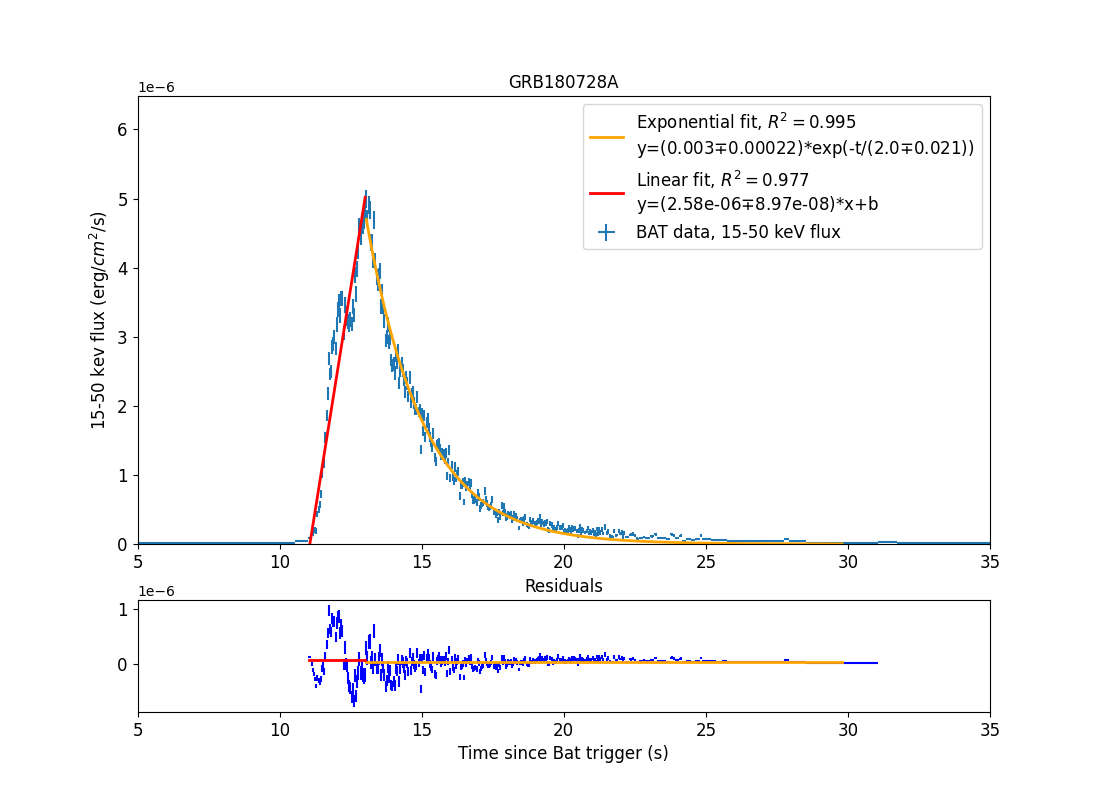}
\caption{Characteristic BAT/Swift 15-50\ keV lightcurve manifesting a sharp linear growth followed by a longer exponential decay. In both curves, the horizontal axis is time since BAT trigger in seconds.}
\label{figurelc}
\end{figure}
\begin{table}
\caption{List of BAT prompt emission GRB lightcurves with linear growth of duration $\Delta$ followed by exponential decay with time constant $\tau$. The $R^2$ values characterize the goodness of the two respective fits (the closer $R^2$ is to unity, the better the fit). Redshift ($z$) values are also shown whenever available.}
\begin{center}
\begin{tabular}{lccccc} 
 %\hline
Name & $\Delta$ & $R^2$ & $\tau$ & $R^2$ & $z$ \\ 
 & & linear & (s) & exponential &\\
  \hline
  \\
 060912A	& 	0.48 &0.867	& 1.6	& 	0.957	& 0.94 \\
070508 & 		0.64 &0.852	&	0.69 & 	0.91	 & 0.82 \\
070917	 & 	0.4 &0.966	&	2.7	& 	0.97	 & - \\
071010B	& 	0.28 &0.846	&	7.5	& 	0.938 &	0.947 \\
080413B	& 	1.14 &0.711	&	1.1	& 	0.966	& 1.1 \\
080727B	& 	0.48 &0.792	&	1.5	& 0.952	& - \\
090201	 & 	0.86 &0.933	&	2.0	&	0.948	& 2.1 \\
091018 & 		1.22 &0.927	&	1.7	&	0.994	& 0.971 \\
091208B	& 	0.44 &0.965	&	0.98	 &	0.952	& 1.063 \\
101024A	& 	0.32 &0.805	&	0.42	 &	0.953	& - \\
110503A	& 	0.772 &0.883	&	2.9	&	0.984	& 1.613 \\
110715A	 & 	3.5 &0.963	&	1.0	&	0.989 &	0.82 \\
111121A	 & 	0.408 &0.83	 &	0.34	 &	0.936	& - \\
120326A	& 	5.428 &0.925	&	5.0	& 	0.944	& 1.798 \\
130831A	& 	2.9 &0.896	& 	4.3	&	0.957	& 0.479 \\
140206A1	& 	0.3 &0.843	&	0.87	& 0.776 &	2.73 \\
140206A2	& 	0.404 &0.875	&	0.74	 &	0.909	& 2.73 \\
140209A	& 	0.256 &0.924	&	0.54	 &	0.941	& - \\
140213A	& 	1.524 &0.879	&	3.2	&	0.965	& 1.208 \\
141004A	& 	0.222 &0.988	&	0.68	 &	0.938	& - \\
150817A	& 	1 &0.903	&	0.44	 &	0.858	& - \\
160425A	& 	1.22 &0.97	 &	0.49	 &	0.773	& 0.555 \\
161001A	& 	0.192 &0.992	&	0.5	& 0.954	& - \\
170711A	& 	1.56 &0.966	&	2.7	& 0.981	& - \\
180728A	& 	1.76 &0.977	&	2.0	& 0.995	& 0.117 \\
200922A	& 	0.748 &0.934	&	3.6	& 0.978	& - \\
201013A	& 	0.88 &0.937	&	0.92	 & 	0.857	& - \\
210306A	& 	0.16 &1	 &	0.72	 & 0.993	& - \\
210514A	& 	0.9 &0.966	&	1.6	& 0.962	& - \\
%\hline
\end{tabular}
\end{center}
\end{table}

\section{Electromagnetic core spindown}

In what follows, we will propose a simple model of electromagnetic core spindown that fits certain observations of the $\gamma$-ray prompt emission in long GRBs. The reason we are doing this is that, based on such a simple model, we can calculate the potential emission of gravitational waves (GW) from this source, and we can propose a particular waveform that may be retreivable from LIGO archival GW data.

Let us consider a GRB stellar precursor of mass $M\gg M_\odot$, and radius $r\gg r_\odot$, where $M_\odot=2\times 10^{33}\ {\rm g}$ and $r_\odot \sim 7\times 10^{10}\ {\rm cm}$ are the solar mass and radius respectively. Let us also assume that the star slowly rotates with angular velocity $\omega_o$ much lower than its breakup velocity (figure~2, left). We will now formulate a simple testable model for core collapse that is followed by electromagnetic spindown and the emission of a GRB jet.

Let us assume that, at $t=0$, the central part of mass $M_c\sim M_\odot$ and radius $r_o\sim r_\odot$ is threaded with magnetic flux $\Psi_c$ and collapses into a compact core of size $r_c\ll r_o$. Angular momentum conservation requires that the initial angular velocity of the core will be equal to
\begin{equation}
\omega_{co}\equiv \omega_c(t=0)=\omega_o \left(\frac{r_o}{r_c}\right)^2 \gg \omega_o\ .
\end{equation}
At all subsequent times, the rotational kinetic energy of the collapsed core is roughly given by
\begin{equation}
E_c\approx \frac{1}{5}M_c \omega_c^2 r_c^2\ .
\label{Ecore}
\end{equation}
As the core collapses, it stretches the magnetic field that it carries with it into a split monopole configuration\footnote{This is a configuration with radial magnetic field pointing away from the center in one hemisphere, and towards the center in the other. The field discontinuity in the equator is supported by equatorial electric currents. Obviously, a split monopole can exist only in a plasma, and not in vacuum.}
 (figure~2, right) with a radial magnetic field equal to
\begin{equation}
B_{rc}=\frac{\Psi_c}{2\pi r^2_c}
\label{Br}
\end{equation}
on the surface of the core\footnote{Notice that if the core is a black hole, it cannot hold the split monopole configuration on its own. In that case too, though, a disk of matter may form in the equatorial region outside the black hole horizon that may hold such a configuration.}. The magnetic field remains also attached to the outer layers of the star at distances $r\gsim r_o$ that did not collapse and rotate much slower than the core. Notice that it is not inconceivable to expect that the internal cavity will not collapse before the light crossing time $r_o/c$. Meanwhile, the magnetic field is wound linearly with time, and an azimuthal magnetic field component grows as 
\begin{equation}
B_{\phi c}(t;\theta)=B_{rc}\int_0^t \frac{\omega_c(t) r_c \sin\theta\ {\rm d}t}{r_o}=\frac{\Psi_c \sin\theta}{2\pi r_c r_o}\int_0^t \omega_c(t) \ {\rm d}t
\label{Bphi}
\end{equation}
(figure~3, left).
%This generates a large scale distributed electric current that threads the split-monopole magnetosphere from the pole ($\theta=0$) to the equator ($\theta=\pi/2$) of the core as 
%\begin{equation}
%I(t;\theta)\equiv 2B_{\phi c}(t;\theta) r_c c \sin\theta =
%\frac{\Psi_c c \sin^2\theta }{\pi r_o}\int_0^t \omega_c(t) \ {\rm d}t\ .
%\label{I}
%\end{equation}
This leads to the electromagnetic spindown of the collapsed core as
\begin{eqnarray}
\dot{E}_{\rm EM}(t) & \approx &
% -\frac{2}{c}\int_0^{\Psi_c} (I(\pi/2)-I(\theta)) \omega_c B_{rc} r_c^2 \sin\theta\ {\rm d}\theta\nonumber\\
%&\approx&
%-\frac{\Psi_c^2 \omega_c }{\pi^2 r_o} \int_0^{\pi/2}\cos^2\theta \sin\theta\ {\rm d}\theta \int_0^t \omega_c(t) \ {\rm d}t\nonumber\\
%&\approx&
%-\frac{\Psi_c^2 \omega_c }{3\pi^2 r_o} \int_0^t \omega_c(t) \ {\rm d}t
-2\times \frac{c}{4\pi}\int_0^{\pi/2}
E_{\theta c}B_{\phi c}2\pi r_c^2 \sin\theta\ {\rm d}\theta\nonumber\\
& \approx &
-c\int_0^{\pi/2}
B_{r c}\frac{r_c\sin\theta \omega_c}{c}B_{\phi c}2\pi r_c^2 \sin\theta\ {\rm d}\theta\nonumber\\
& \approx &
-\frac{\Psi_c^2}{4\pi^2 r_o}\int_0^{\pi/2} \sin^3\theta\ {\rm d}\theta \ \omega_c\int_0^t \omega_c\ {\rm d}t\nonumber\\
& = &
-\frac{\Psi_c^2 }{3\pi^2 r_o}\ \omega_c \int_0^t \omega_c(t) \ {\rm d}t
\label{EdotEM}
\end{eqnarray}
In the above expression, we integrated the radial component of the Poynting flux vector $(c/4\pi)\ {\bf E}\times{\bf B}$ over each hemisphere of the core, and considered the contribution from both hemispheres. 
Differentiating eq.~(\ref{Ecore}) with respect to time and equating it to eq.~(\ref{EdotEM}) we obtain
\begin{equation}
\dot{\omega}_c =-
\frac{5\Psi_c^2}{6\pi^2 M_c r_c^2 r_o}\int_0^t \omega_c(t) \ {\rm d}t\ .
\label{Edot2}
\end{equation}
Diferentiating eq.~(\ref{Edot2}) once again with respect to time we obtain
\begin{equation}
\ddot{\omega}_c=-\frac{1}{\tau^2_{\rm rise}}\omega_c\ ,
\end{equation}
where
\begin{eqnarray}
\tau_{\rm rise} & \equiv & \left(\frac{6\pi^2 M_c r_c^2 r_o}{5\Psi_c^2}\right)^{1/2}\nonumber\\
& \sim & 0.6\ {\rm sec}\ \left(\frac{M_c}{M_\odot}\right)^{1/2}
\left(\frac{B_{rc}}{10^{16}\ {\rm G}}\right)^{-1}
\left(\frac{r_c}{10\ {\rm km}}\right)^{-1}\left(\frac{r_o}{r_\odot}\right)^{1/2}\nonumber\\
\end{eqnarray}
From the above we obtain that, at early times $t\lsim 1$~s following core collapse,
\begin{eqnarray}
\omega_c(t) & \approx & \omega_{co} \cos(\frac{t}{\tau_{\rm rise}})\nonumber\\
&\approx & 
\omega_{co} \label{omegac1}\\
\dot{E}_{\rm EM}(t)&\approx& -\frac{\Psi_c^2\omega_{co}^2}{6\pi^2 r_o}\tau_{\rm rise}\sin(\frac{2t}{\tau_{\rm rise}})\nonumber\\
&\approx&
-\frac{\Psi_c^2\omega_{co}^2}{3\pi^2 r_o}t
%=-\frac{\Psi_c^2 \omega_o^2}{3\pi^2 r_c}t
\label{Edotearly}
\end{eqnarray}
We thus see that, at short timescales, the electromagnetic energy loss of the core grows almost linearly with time with a slope that depends only on the magnetic flux that threads the collapsed core and has been carried by it during the core collapse.

\begin{figure}
 \centering
 %\vspace{3cm}
 \includegraphics[width=9cm,height=5cm,angle=0.0]{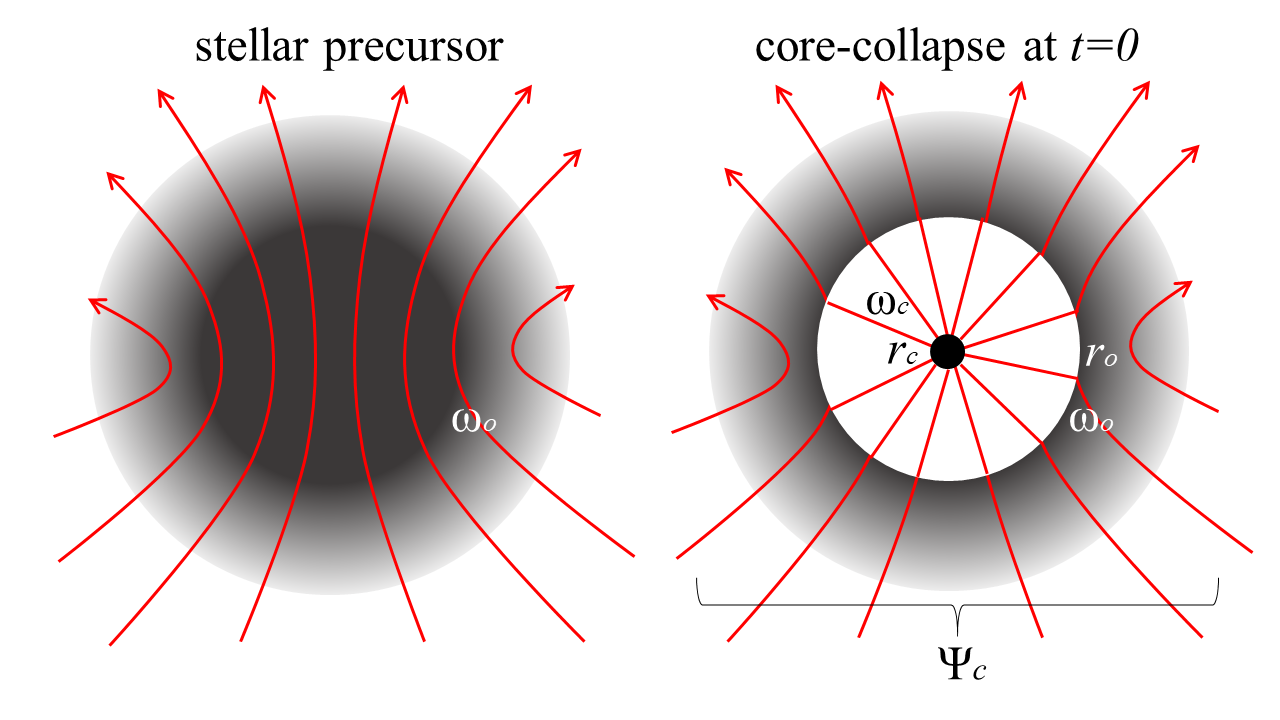}
\caption{Schematic of core collapse inside a high mass slowly rotating magnetized star. The core carries with it the magnetic flux $\Psi_c$ that threads it and forms a split-monopole configuration.}
\label{figure1}
\end{figure}

After some time $\Delta$ on the order of the light crossing time inside the collapsed region of size $r_o$, the wound magnetic field becomes comparable to the radial magnetic field in the core, the electromagnetic field configuration pushes outwards through the overlying layers of the progenitor star, and a GRB jet is driven out of the core (figure~3, right). In that stage of the core evolution, the winding of the field is halted, the distributed magnetospheric electric current may be approximated by the Michel split-monopole expression
\begin{equation}
I_{\rm Michel}(t;\theta)=\frac{\Psi_c \omega_c(t)\sin^2\theta}{\pi}
\label{IGRB}
\end{equation}
\citep{M82}, the azimuthal magnetic field is given by
\begin{equation}
B_{\phi c} \approx \frac{I_{\rm Michel}(t;\theta)}{2r_c c\sin\theta}\ ,
\end{equation}
and the spinning core loses rotational energy at a rate
\begin{eqnarray}
\dot{E}_{\rm GRB}(t) & \approx & 
-2\times \frac{c}{4\pi}\int_0^{\pi/2}
E_{\theta c}B_{\phi c}2\pi r_c^2 \sin\theta\ {\rm d}\theta\nonumber\\
%&\approx& -\frac{2}{c}\int_0^{\Psi_c} (I(\pi/2)-I(\theta)) \omega_c B_{rc} r_c^2 \sin\theta\ {\rm d}\theta\nonumber\\
%&\approx&
%-\frac{\Psi_c^2 \omega^2_c }{2\pi^2 c} \int_0^{\pi/2 c}\cos^2\theta \sin\theta \ {\rm d}\theta\nonumber\\
&\approx&
-\frac{\Psi_c^2 \omega^2_c }{6\pi^2 c}\ .
\label{EdotEMGRB}
\end{eqnarray}
We emphasize once again that the latter expression is approximately valid during the decay phase of the GRB prompt emission. As before, we differentiate eq.~(\ref{Ecore}) with respect to time and equate it to eq.~(\ref{EdotEMGRB}). We then obtain that, at late times following the release of the electromagnetic GRB jet from the stellar interior,
\begin{eqnarray}
\omega_c(t)&\approx&  \omega_{co}\cos(\frac{\Delta}{\tau_{\rm rise}})\ 
%\left(\frac{r_o}{r_c}\right)^{1/2}
{\rm e}^{-\frac{t-\Delta}{2\tau}}
\ ,\label{omegac2}\\
\dot{E}_{{\rm EM}\ {\rm GRB}}(t)&\approx& -\frac{\Psi_c^2 \omega_{co}^2}{6\pi^2 c}\cos^2(\frac{\Delta}{\tau_{\rm rise}})\ 
%\left(\frac{r_c}{r_o}\right)
{\rm e}^{-\frac{t-\Delta}{\tau}}
\ ,
\label{Edotlate}
\end{eqnarray}
where,
\begin{eqnarray}
\tau &\equiv& \frac{6\pi^2 M_c r_c^2 c}{5\Psi^2_c}
%\equiv \frac{\tau_{\rm rise}^2}{r_o/c}\sim \frac{\tau_{\rm rise}^2}{\Delta}
\nonumber\\
&\approx& 0.2\ {\rm sec}\ \left(\frac{M_c}{M_\odot}\right)\left(\frac{B_{rc}}{10^{16}\ {\rm G}}\right)^{-2}\left(\frac{r_c}{10\ {\rm km}}\right)^{-2}
\end{eqnarray}
is the electromagnetic decay timescale \citep{CNP14}. We are not in a position to discuss here what fraction of the energy loss in eqs.~(\ref{Edotearly}) and (\ref{Edotlate}) is channelled into the $\gamma$-rays that we observed with Swift. We can only speculate that, in analogy to pulsars \citep[e.g.][]{Ketal18}, particle acceleration to ultra-relativistic energies and consequent curvature/synchrotron radiation will certainly take place in the current sheet that develops after the formation of the split monopole configuration (see footnote~2).
Alternatively, in the realm of speculation again, one can assume that all high energy photons resulting from the electromagnetic dissipation, are converted, due to the compactness of the source, into ${\rm e}^+{\rm e}^-$ pairs of an extremely rich in pairs plasma that radiates at near 1~MeV as suggested by \citet{P86}. 
It is thus natural to assume that the observed $\gamma$-ray luminosity will follow the same time evolution as the electromagnetic radiation of the collapsed core.

Notice that it is rather common for the prompt GRB emission to manifest several stages of fast linear growth followed by a slow exponential decay. We interpret these as subsequent events of core collapse from regions of size $\sim r_o$. Each such collapse is followed by the emission of an electromagnetic GRB jet that results in the exponential slowing down of the core rotation. Finally, notice that $\tau_{\rm rise}^2=\tau r_o/c\sim \tau \Delta$, or in other words, $\tau_{\rm rise}$ is comparable to $(\tau \Delta)^{1/2}$.
%Eqs.~(\ref{EdotEM}) and (\ref{EdotEMGRB}) give comparable values at times $\Delta\sim r_o/2c$ from each core collapse event. This allows us to estimate the size of the stellar interior that collapsed into the central fast spinning core. 

\begin{figure}
 \centering
% \vspace{3cm}
 \includegraphics[width=9cm,height=5cm,angle=0.0]{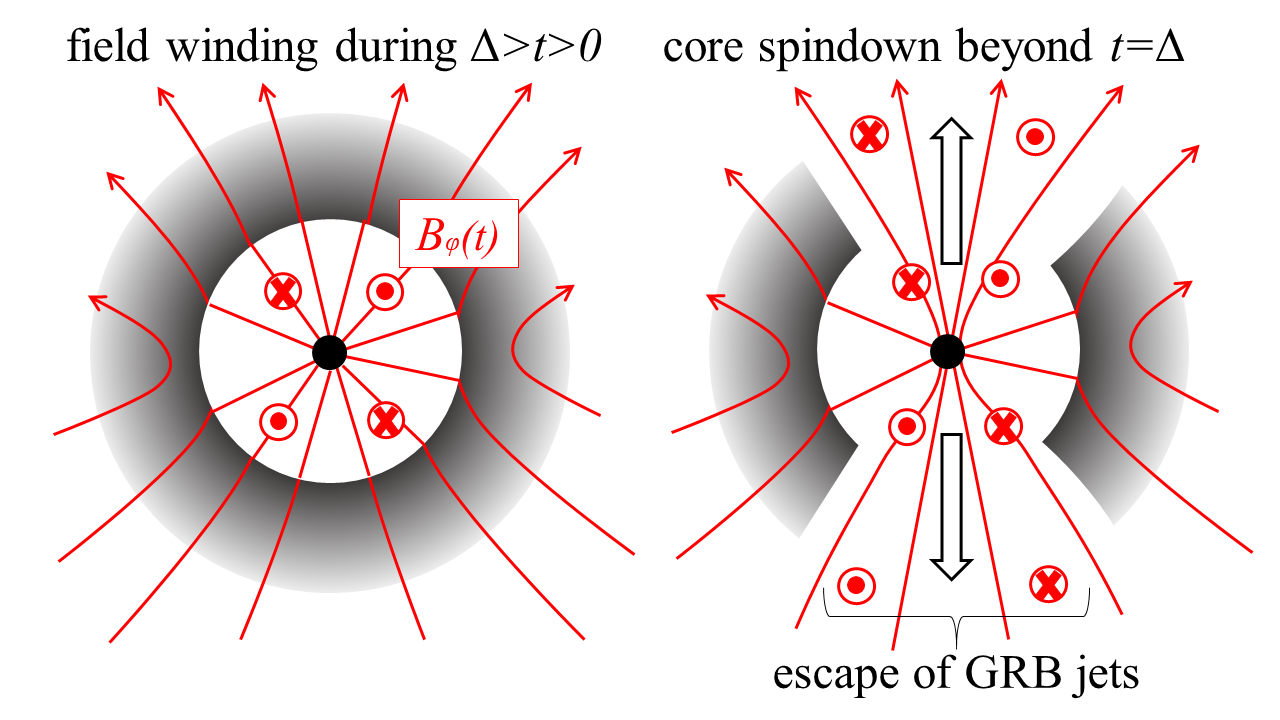}
\caption{Following core collapse, the magnetic field is wound by the fast core rotation. The azimuthal magnetic field and the electromagnetic radiation both grow linearly with time up to a time $\Delta$ that the electromagnetic field escapes through the overlying stellar layers in the form of a GRB jet. The system may experience several such events that will appear as subsequent bursts in the GRB prompt emission. Beyond that time, open field lines are not wound any further, and the electromagnetic radiation decreases exponentially with time. Field winding continues along field lines that connect the core to the surrounding stellar layer.}
\label{figure2}
\end{figure}

\section{GW from the spinning core}

Up to now, all calculations have been performed under the assumption of axial symmetry. Under this assumption, no GW are expected from this system. However, if we are willing to break the axial symmetry, then all events singled out in Table~1 are potential sources of GW that may be observable with a GW Observatory. In fact the core may form non-axisymmetrically, e.g. consist of a binary system of compact cores, or manifest growing non-axisymmetric perturbations. In that case, a blind search for GWs in archival LIGO data should consider the frequency evolution described by eqs.~(\ref{omegac1}) \& (\ref{omegac2}), namely
\begin{eqnarray}
\omega_c(t) & = & \omega_c(t=\Delta) \times 
\left\{\begin{array}{lr}
        \frac{\cos(t/\tau_{\rm rise})}{\cos(\Delta/\tau_{\rm rise})} & \text{for  } 0\leq t\leq \Delta\\ \\
        {\rm e}^{-\frac{t-\Delta}{2\tau}} & \text{for  } t> \Delta
        \end{array}\right. \nonumber\\
        \label{downchirp}
\end{eqnarray}
The search should be performed during the particular prompt emission events of Table~2 that took place when the LIGO GW observatory was operational.  We have no estimate of the expected GW strain. Nevertheless, we urge the LIGO Team to look into their archival data during those particular time intervals to test our simple hypothesis. We emphasize that, GWs from such a system, if observed, will look like a `down-chirp' (as opposed to the `up-chirp' or just `chirp' recorded during binary compact star mergers).

One (less hopeful as we will see) factor that may break axial symmetry during the core collapse and subsequent field-winding phase is a non-axisymmetric tilted magnetic field component $B_{\rm na}$ that is wound into an azimuthal component $B_{{\rm na}\ \phi}$. This will deform the core into an $m=2$ non-axisymmetric mode that will depend on the magnitude of $B_{{\rm na}\ \phi}^2$. We will now check whether this effect is observable with present-day GW detectors. We may estimate the evolution of the non-axisymmetric wound field component as
\begin{eqnarray}
B_{{\rm na}\ \phi}(t) & \sim & B_{\rm na}(t=0)\frac{r_c}{r_o}\int_0^t \omega_c(t)\ {\rm d}t\nonumber\\
& = & B_{{\rm na}\ \phi}(t=\Delta) \times \left\{\begin{array}{lr}
        \frac{\sin(t/\tau_{\rm rise})}{\sin(\Delta/\tau_{\rm rise})} & \text{for  } 0\leq t\leq \Delta\\ \\
        1 + \frac{2\tau}{\tau_{\rm rise}} \left[ 1-{\rm e}^{-\frac{t-\Delta}{2\tau}}\right] & \text{for  } t> \Delta
        \end{array}\right. \nonumber\\
        \label{Bna}
\end{eqnarray}
%Because we refer here to the non-axisymmetric field component, its azimuthal dependence will be of the form $B_{\rm na}(\phi)\sim \cos\phi$, and therefore, the azimuthal dependence of $B_{\rm na}^2$ will be of the form $B_{\rm na}^2\sim \cos(2\phi)$ (m=2 dependence).
%The radial deformation $\delta r_c$ of the core resulting from such a non-axisymmetric distribution of the above field may be crudely estimated as
%\begin{eqnarray}
%\delta r_c 
%& \sim & \frac{B_{\rm na}^2 r_c^3}{GM_c^2/r_c} r_c
%%\sim \frac{B_{\rm na}^2G^3 M_c^2}{c^8}
%\nonumber\\
%& \sim & 10^{3}\left(\frac{B_{\rm na}(t=\Delta)}{10^{16}\ {\rm G}}\right)^2
%\left(\frac{\Delta}{1\ {\rm s}}\right)^2 {\rm cm}\nonumber\\
%& & \times
%\left\{\begin{array}{lr}
 %       \left(\frac{t}{\Delta}\right)^2 & \text{for  } 0\leq t\leq \Delta\\
%        \left(1 + \frac{2\tau}{\Delta} \left[ 1-{\rm e}^{-\frac{t-\Delta}{2\tau}}\right]\right)^2 & \text{for  } t> \Delta
%        \end{array}\right. \nonumber\\
%        \label{waveform}
%%\end{eqnarray}
%%, $\Psi_\odot\equiv 2\pi r_o^2 \times 1\ {\rm G}$, and $r_o\equiv r_\odot=7\times 10^{10}\ {\rm cm}$.
%In the above numerical estimate we took $M_c\sim M_\odot=2\times 10^{33}\ {\rm g}$, and $r_c\sim 10\ {\rm km}$.
As we said before, the azimuthal dependence of $B_{{\rm na}\ \phi}^2$ will introduce an $m=2$ non-axisymmetric gravitational perturbation $\sim \cos(2\phi)$. We may estimate the strain $h$ measured on earth due to the gravitational radiation signal by considering the equivalent `mass' $(B^2_{{\rm na}\ \phi}/8\pi c^2)(2\pi r_c^2\times 2r_c)$ of each hemisphere filled with the non-axisymmetric magnetic field energy density corrotating with the core at some distance $\sim r_c$ above the core's surface, namely 
\begin{eqnarray}
h 
& \sim & \frac{G B_{\rm na}^2 r_c^3}{c^4}\times
\frac{1}{r}\times  \frac{\omega_c^2 r_c^2}{c^2}
\nonumber\\
 & \sim & 3\times 10^{-24}\left(\frac{{B_{{\rm na}\ \phi}\ \phi}(t=\Delta)}{10^{17}\ {\rm G}}\right)^2
\left(\frac{\omega_{c}(t=\Delta)}{10^4\ {\rm rad}/{\rm s}}\right)^2
\left(\frac{r}{10\ {\rm Mpc}}\right)^{-1}\nonumber\\
& & \times
\left\{\begin{array}{lr}
        \left(\frac{\sin(2t/\tau_{\rm rise})}{\sin(2\Delta/\tau_{\rm rise})}\right)^2 & \text{for  } 0\leq t\leq \Delta\\ \\
        \left(1 + \frac{2\tau}{\tau_{\rm rise}} \left[ 1-{\rm e}^{-\frac{t-\Delta}{2\tau}}\right]\right)^2 {\rm e}^{-\frac{t-\Delta}{\tau}} & \text{for  } t> \Delta
        \end{array}\right. \nonumber\\
        \label{waveform}
\end{eqnarray}
where $r$ is the distance from the GRB to earth. This value may be borderline observable with the Einstein Telescope in extremely magnetized kHz core-collapse events with fiducial values $B_{{\rm na}\ \phi}(t=\Delta)\sim 10^{17}\ {\rm G}$ and $\omega_{c}(t=\Delta)=10^4\ {\rm rad}/{\rm s}$ within a distance of about 10~Mpc around the earth (see figure~4). The waveform described by eq.~(\ref{waveform}) is shown in figure~5 for fiducial values $\omega_{co}/(2\pi)=30\ {\rm Hz}$, $\Delta=0.3$, $\tau=1\ {\rm s}$, and $\tau_{\rm rise}=0.6\ {\rm s}$ (we chose such a low frequency only to plot the GW waveform). 
\begin{figure}
 \centering
 \includegraphics[width=9cm,height=11cm,angle=0.0]{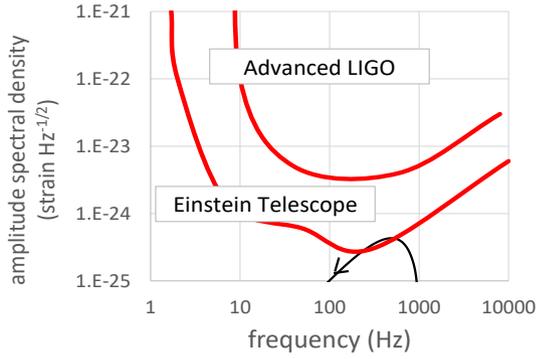}
 \vspace{-3cm}
\caption{Thin black line: fiducial sensitivity curve starting from $\omega_{co}=2\pi\times 1000\ {\rm rad}/{\rm s}$. The evolution of this event is traced from the lower right to the upper left corner of the curve (a GW `down-chirp'). Here also, $\Delta=0.3\ {\rm s}$, $\tau=1\ {\rm s}$, and $\tau_{\rm rise}=0.6\ {\rm s}$. Thick red lines: approximate Advanced~LIGO and Einstein Telescope sensitivity limits. The events we are proposing are probably borderline unobservable.}
\label{figure4}
\end{figure}
\begin{figure}
 \centering
 \includegraphics[width=9cm,height=12cm,angle=0.0]{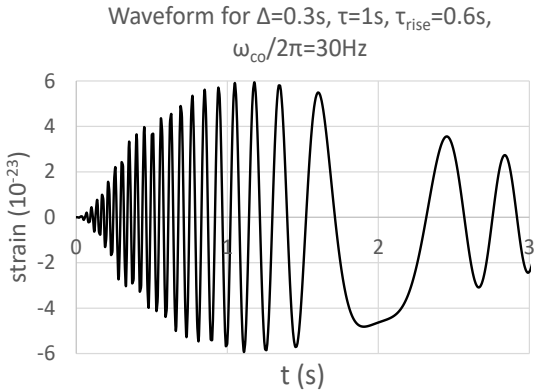}
 \vspace{-3cm}
\caption{Predicted GW waveform according to the parametric formulation of eqs.~(\ref{waveform}) and the parameter values shown on the plot.}
\label{figure5}
\end{figure}

In conclusion, the waveform shown in figure~5 manifests two effects that take place during the evolution of the core collapse according to our simple electrodynamic model. The first one is the growth of the gravitational perturbation due to the continuous winding of the non-axisymmetric magnetic field. Even if there is no such field component, the core may still be non-axisymmetric or manifest growing non-axisymmetric perturbations that emit gravitational waves. The second one is the exponential electrodynamic spindown of the core which takes place with or without a non-axisymmetric magnetic field component.

\section*{Data Availability}

The data underlying this article will be shared on reasonable request to the corresponding author.

\begin{table}
\caption{GRBs with linear growth followed by exponential decay in their prompt emission that occured during the operation of the LIGO GW Observatory. We urge the LIGO Team to look for the particular GW waveform described by eqs.~(\ref{downchirp}) \& (\ref{waveform}) during the time intervals indicated in this Table.}

\begin{center}
\begin{tabular}{ lccccc} 
 %\hline
Name &  Date & Start of &  End of  \\ 
 & &  linear growth & exponential decay  \\
 & & (UTC)  & (UTC)  \\
  \hline
\\
 160425A	&  April 25, 2016 &23:26:10.961			 &	 	 23:26:12.661		 \\
161001A	& October 1, 2016 	 & 01:05:18.12		&		01:05:19.18 	 \\
170711A	& July 11, 2017	& 22:20:24.432 	&		22:20:35.526 	 \\
200922A	& September 22, 2020	& 12:06:45.441 	&		12:06:59.281 	 \\
201013A	& 	October 13, 2020 & 03:46:30.224 	&	 03:46:34.884  \\
210306A	& 	March 6, 2021 & 03:54:04.682		&	 03:54:09.502  \\
210514A	& 	May 14, 2021 & 18:23:59.102		&		18:24:06.162  \\
%\hline
\end{tabular}
\end{center}
\end{table}

%\section*{Acknowledgements}
%
%We thank 

%$$$$$$$$$$$$$$$$$$$$$$$$$$$$$$$

\bibliographystyle{mn2e}

\newpage
\begin{figure}
 \centering
 %\vspace{3cm}
 \includegraphics[width=9cm,height=7cm,angle=0.0]{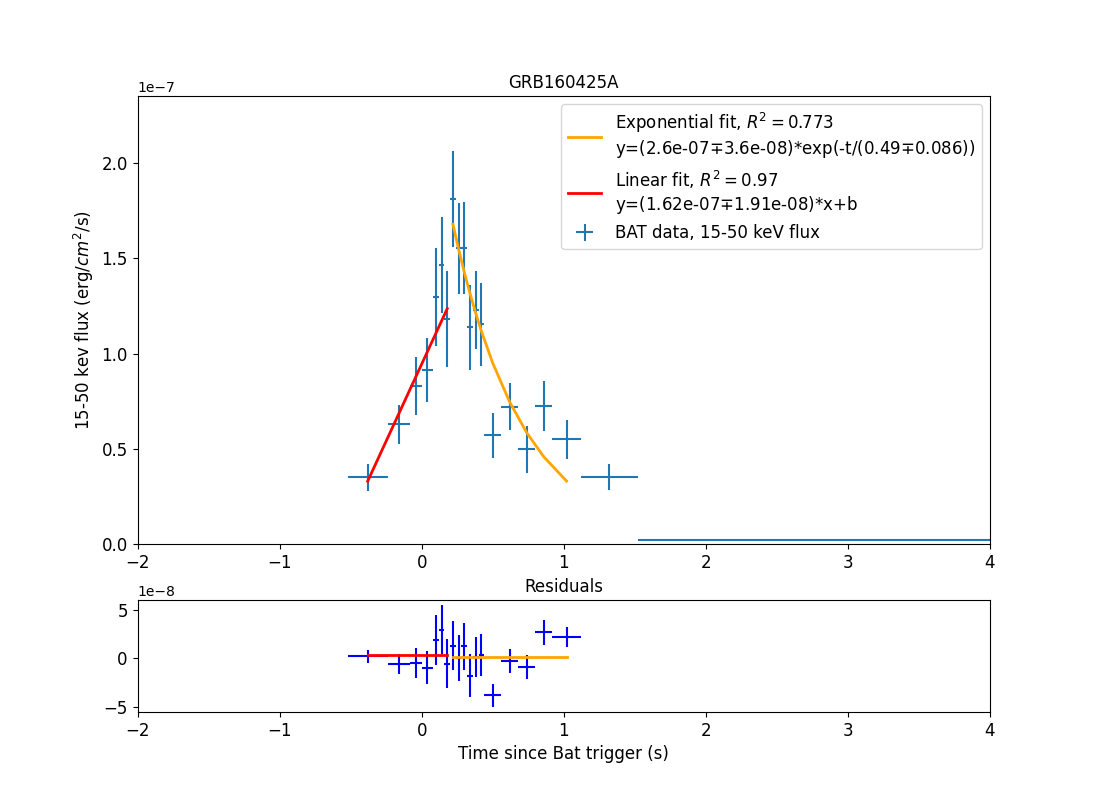}
 \includegraphics[width=9cm,height=7cm,angle=0.0]{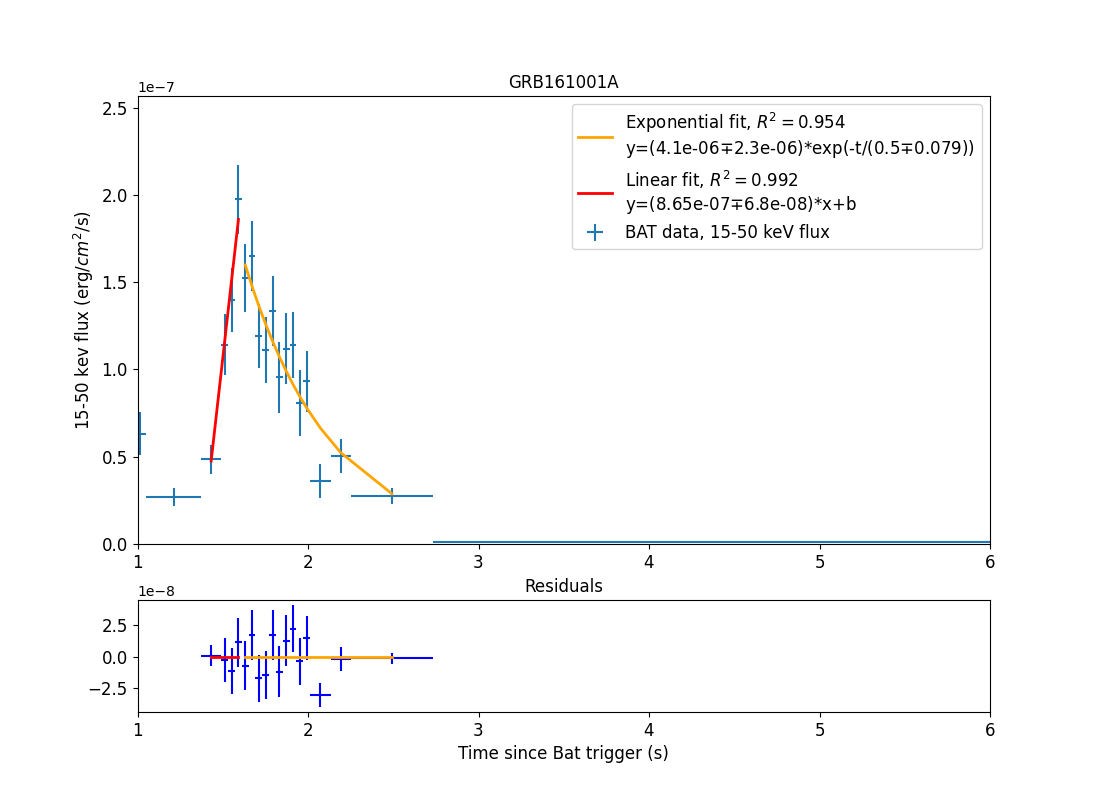}
 \includegraphics[width=9cm,height=7cm,angle=0.0]{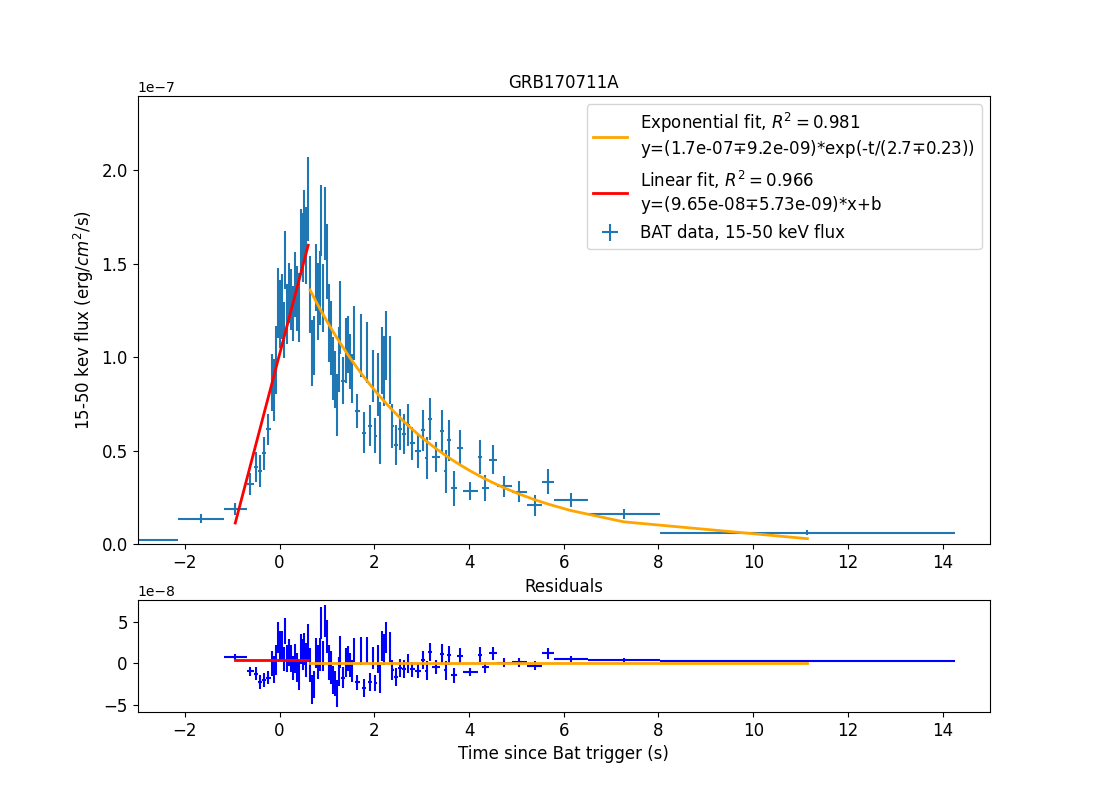}
 \end{figure}
\begin{figure}
 \centering
 \includegraphics[width=9cm,height=7cm,angle=0.0]{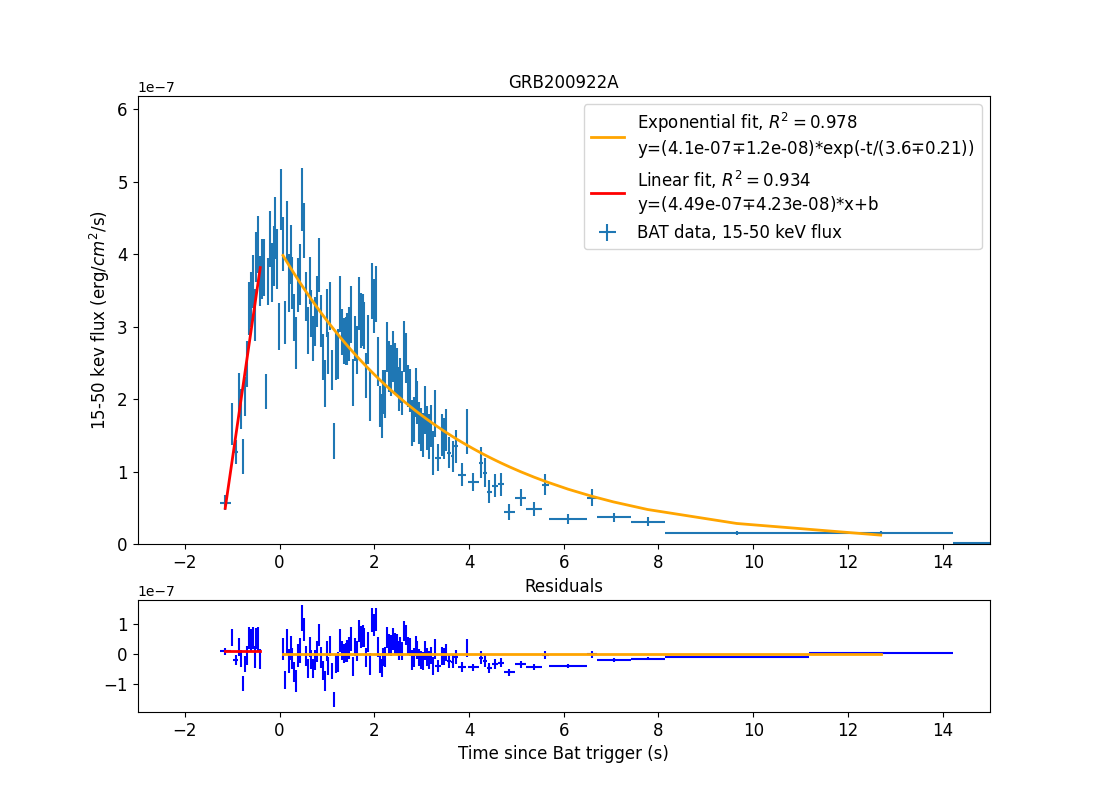}
 \includegraphics[width=9cm,height=7cm,angle=0.0]{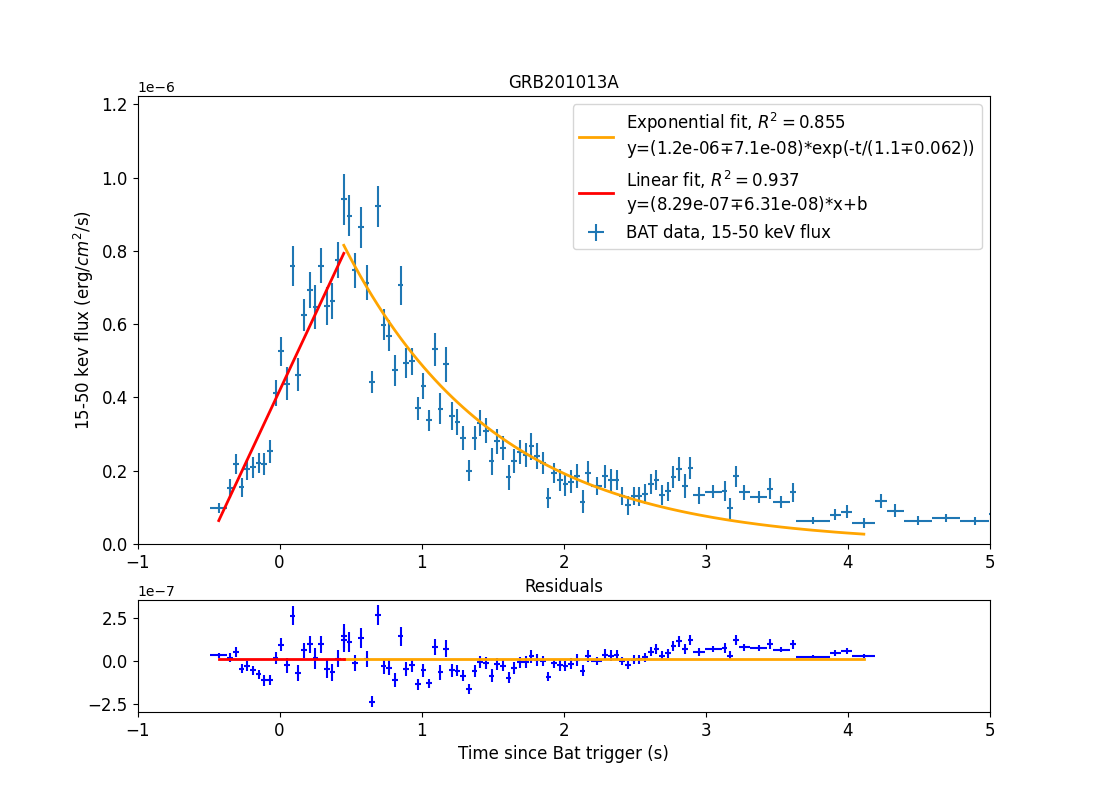}
 \includegraphics[width=9cm,height=7cm,angle=0.0]{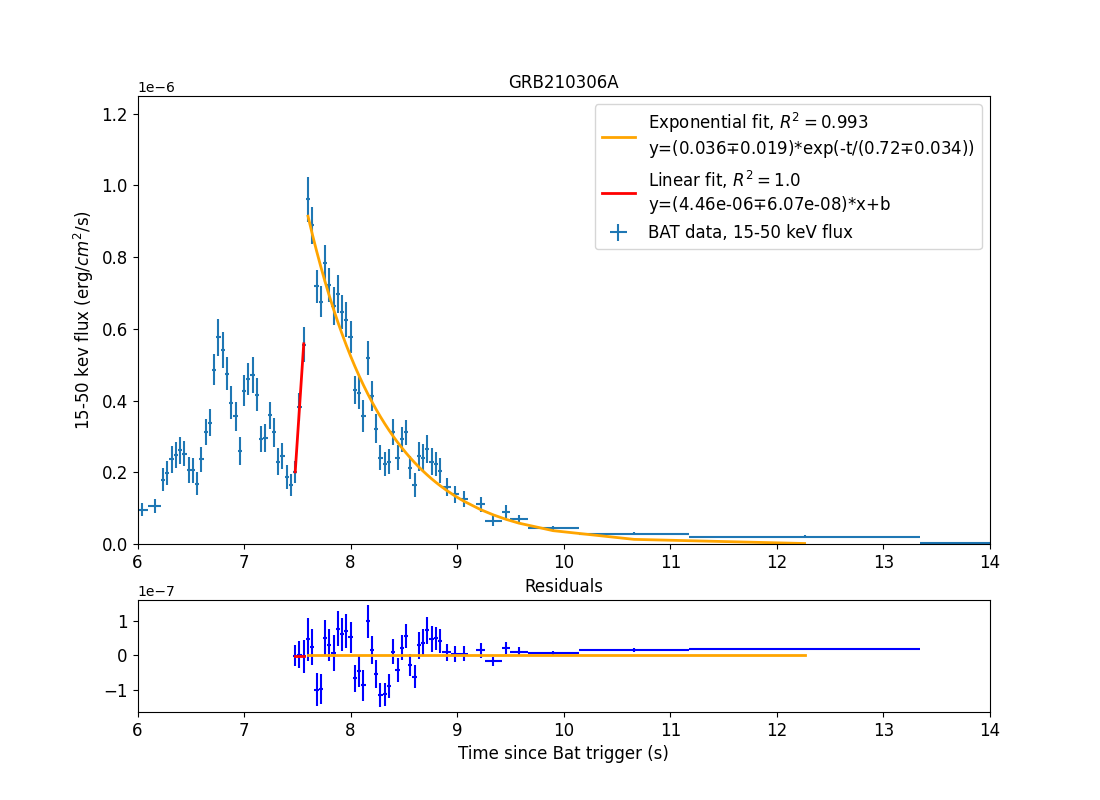}
 \end{figure}
\begin{figure}
 \centering
 \includegraphics[width=9cm,height=7cm,angle=0.0]{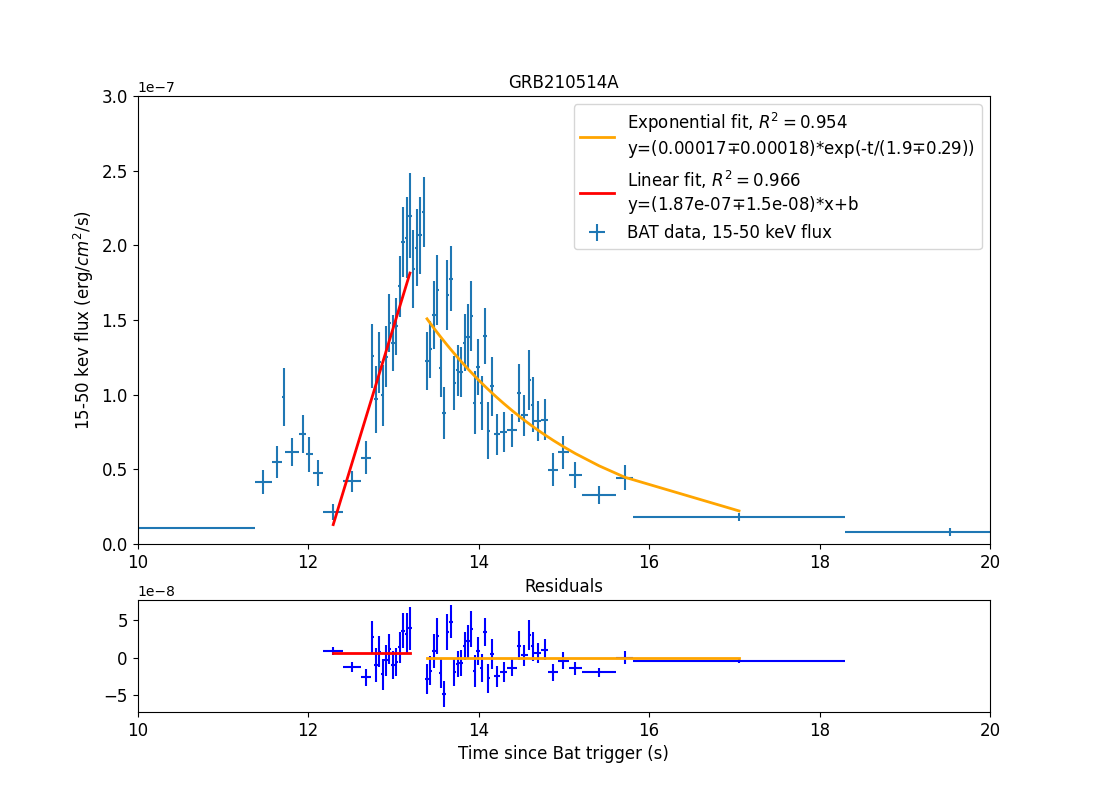}
 \caption{List of the characteristic BAT light curves from the potential GW sources of Table~2.}
\label{figure5}
\end{figure}

\end{document}